\documentclass[11pt]{article}
\usepackage[dvips]{graphicx}
\usepackage{psfrag}
\usepackage{amssymb}
\usepackage{amsmath}
\usepackage{amscd}
\usepackage{latexsym}
\usepackage{bbm}

\catcode`@=11
\makeatletter\renewcommand{\section}{\@startsection
{section}{1}{\z@}{-3.5ex plus -1ex minus
    -.2ex}{2.3ex plus .2ex}{\large\bf }}

\makeatletter\renewcommand{\subsection}{\@startsection{subsection}{2}{\z@}{-3.25ex
plus -1ex minus
   -.2ex}{1.5ex plus .2ex}{\bf }}

\numberwithin{equation}{section}
\newcounter{saveeqn}

\catcode`@=12

\def\a{\alpha}
\def\b{\beta}
\def\ga{\gamma}
\def\la{\lambda}

\def\de{\delta}
\def\eps{\epsilon}
\def\ve{\varepsilon}

\def\th{\theta}

\def\p{\phi}
\def\vp{\varphi}
\def\om{\omega}
\def\O{\Omega}
\newcommand{\unity}{\mathbbm{1}}
\newcommand{\C}{\mathbb C}
\newcommand{\R}{\mathbb R}

\newcommand{\Gbb}{\mathbb G}
\newcommand{\Jcal}{{\cal J}}
\newcommand{\Gcal}{{\cal G}}
\newcommand{\Acal}{{\cal A}}
\newcommand{\Lcal}{{\cal L}}
\newcommand{\Mcal}{{\cal M}}
\newcommand{\Fcal}{{\cal F}}
\newcommand{\Ncal}{{\cal N}}

\newcommand{\Pcal}{{\cal P}}

\newcommand{\gfrak}{{\mathfrak g}}

\newcommand{\sh}{{\hat{\smash{\sigma}}}}
\newcommand{\mh}{{\hat{\smash{\mu}}}}
\newcommand{\nh}{{\hat{\smash{\nu}}}}

\newcommand{\lh}{{\hat{\lambda}}}
\newcommand{\lrc}{\mathop{\lrcorner}}
\newcommand{\hra}{\mathop{\hookrightarrow}}
\def\1{{\bar 1}}
\def\im{\textrm{i}}
\def\tr{\textrm{tr}}
\def\diff{\textrm{d}}
\def\pa{\mbox{$\partial$}}
\def\sfrac#1#2{{\textstyle\frac{#1}{#2}}}
\def\+{\dagger}
\def\={\ =\ }
\def\und{\qquad\textrm{and}\qquad}
\def\and{\quad\textrm{and}\quad}
\def\with{\quad\textrm{with}\quad}
\def\for{\quad\textrm{for}\quad}
\def\or{\quad\textrm{or}\quad}
\def\Id{\mathrm{Id}}

\begin{document}

\begin{titlepage}
\setcounter{page}{0}
\begin{flushright}
ITP--UH--21/14
\end{flushright}

\hspace{2.0cm}

\begin{center}

{\Large\bf
Sigma-model limit of Yang-Mills instantons in higher dimensions
}

\vspace{12mm}

{\large  Andreas Deser, Olaf Lechtenfeld${}^{\times}$ and Alexander D. Popov
}\\[8mm]

\noindent {\em
Institut f\"ur Theoretische Physik\\
Leibniz Universit\"at Hannover \\
Appelstra\ss e 2, 30167 Hannover, Germany }\smallskip\\
{Email: Andreas.Deser, Olaf.Lechtenfeld, Alexander.Popov@itp.uni-hannover.de}\\[6mm]

\noindent ${}^{\times}${\em
Riemann Center for Geometry and Physics\\
Leibniz Universit\"at Hannover \\
Appelstra\ss e 2, 30167 Hannover, Germany }\\[6mm]

\vspace{10mm}

\begin{abstract}
\noindent
We consider the Hermitian Yang-Mills (instanton) equations for connections on vector bundles
over a $2n$-dimensional K\"ahler manifold $X$ which is a product $Y\times Z$ of $p$- and $q$-dimensional
Riemannian manifold $Y$ and $Z$ with $p+q=2n$. We show that in the adiabatic limit, when the metric
in the $Z$ direction is scaled down, the gauge instanton equations on $Y\times Z$ become sigma-model
instanton equations for maps from $Y$ to the moduli space $\Mcal$ (target space) of gauge instantons on $Z$
if $q\ge 4$. For $q<4$ we get maps from $Y$ to the moduli space  $\Mcal$ of flat connections on $Z$. Thus,
the Yang-Mills instantons on $Y\times Z$ converge to sigma-model instantons on $Y$ while $Z$ shrinks to a point.
Put differently, for small volume of $Z$, sigma-model instantons on $Y$ with target space  $\Mcal$ approximate
Yang-Mills instantons on $Y\times Z$.
\end{abstract}

\end{center}
\end{titlepage}

\section{Introduction and summary}

The Yang-Mills equations in two, three and four dimensions were intensively studied both in physics and mathematics.
In mathematics, this study  (e.g.\ projectively flat unitary connections and stable bundles in $d=2$~\cite{1}, the
Chern-Simons model and knot theory in $d=3$, instantons and Donaldson invariants~\cite{2} in $d=4$ dimensions) has
yielded a lot of new results in differential and algebraic geometry. There are also various interrelations between gauge
theories in two, three and four dimensions. In particular, Chern-Simons theory in  $d=3$ dimensions reduces
to the theory of flat connections in $d=2$ (see e.g.~\cite{3,4}). On the other hand, the gradient flow equations for
Chern-Simons theory on a $d=3$ manifold $Y$ are the first-order anti-self-duality equations on $Y\times\R$, which play a
crucial role in $d=4$ gauge theory.

The program of extending familiar constructions in gauge theory, associated to problems in low-dimensional topology,
to higher dimensions was proposed by Donaldson and Thomas in the seminal
paper~\cite{5} (see also~\cite{6}) and developed in~\cite{7}-\cite{13} among others. An important role in this investigation
is played by first-order gauge-field equations which  are a generalization of the anti-self-duality equations in $d=4$ to
higher-dimensional manifolds with special holonomy (or, more generally, with $G$-structure~\cite{14, 15}). Such equations were
first introduced in~\cite{16} and further considered in~\cite{17}-\cite{21} (see also references therein).

Instanton equations  on a $d$-dimensional Riemannian manifold $X$ can be introduced  as follows~\cite{16, 5, 9}.
Suppose there exist  a 4-form $Q$ on $X$. Then there exists a $(d{-}4)$-form $\Sigma:=*Q$, where $*$ is the Hodge operator on $X$.
Let $\Acal$ be a connection on a bundle $E$ over $X$ with curvature $\Fcal =\diff\Acal + \Acal\wedge\Acal$. The
generalized anti-self-duality (instanton) equation on the gauge field then is~\cite{9}
\begin{equation}\label{1.1}
 *\Fcal + \Sigma\wedge\Fcal \=0\ .
\end{equation}
For $d>4$ these equations can be defined on manifolds $X$ with {\it special holonomy}, i.e.\ such that the holonomy group $G$
of the Levi-Civita connection on the tangent bundle $TX$ is a subgroup in SO$(d)$. Solutions of (\ref{1.1}) satisfy the
Yang-Mills equation
\begin{equation}\label{1.2}
\diff \, *\Fcal + \Acal\wedge *\Fcal - (-1)^d *\Fcal\wedge\Acal \=0\ .
 \end{equation}
The instanton equation  (\ref{1.1}) is also well defined on manifolds $X$ with non-integrable $G$-{\it structures}, i.e.\
when $\diff\Sigma\ne 0$. In this case  (\ref{1.1}) implies the Yang-Mills equation with (3-form) torsion
$T:=*\diff\Sigma$, as is discussed e.g.\ in~\cite{22}-\cite{25}.

Manifolds $X$ with a $(d{-}4)$-form $\Sigma$ which admits the instanton equation (\ref{1.1}) are usually {\it calibrated\/} manifolds
with {\it calibrated submanifolds}. Recall that a calibrated manifold is a Riemannian manifold $(X, g)$ equipped with a closed $p$-form $\vp$ such
that for any oriented $p$-dimensional subspace $\zeta$ of $T_xX$, $\vp\!\mid_\zeta\,\le vol_\zeta$ for any  $x\in X$, where $vol_\zeta$ is the volume
of $\zeta$ with respect to the metric $g$~\cite{26}. A $p$-dimensional submanifold $Y$ of $X$ is said to be a calibrated submanifold
with respect to $\vp$ ($\vp$-calibrated) if $\vp\!\mid_Y\,= vol_Y$~\cite{26}. In particular, suitably normalized powers of the K\"ahler form on
a K\"ahler manifold are calibrations, and the calibrated submanifolds are complex submanifolds. On a $G_2$-manifold one has a 3-form which
defines a calibration, and on a Spin(7)-manifold the defining 4-form (the Cayley form) is a calibration as well~\cite{5,6}.

It is not easy to construct solutions of (\ref{1.1}) for $d>4$ and to describe their moduli space.\footnote{Some explicit solutions for particular
manifolds $X$ were constructed e.g.\ in~\cite{20, 22, 24, 13, 25}.}
It was shown by Donaldson, Thomas, Tian~\cite{5,9} and others that the
{\it adiabatic limit\/} method provides a useful and powerful tool.
The adiabatic limit refers to the geometric process of shrinking a metric in some directions
while leaving it fixed in the others.\footnote{
In lower dimensions, the adiabatic limit was successfully used for a description of solutions
to the $d{=}2{+}1$ Ginzburg-Landau equations and to the $d{=}4$ Seiberg-Witten monopole equations
(see e.g.~reviews~\cite{ser1,ser2} and references therein).}
It is assumed that on $X$ there is a family $\Sigma_{\ve}$ of $(d{-}4)$-forms with a real parameter
$\ve$ such that $\Sigma_0=\lim\limits_{\ve\to 0}^{}\Sigma_\ve$ defines a calibrated submanifold $Y$ of $X$. Then one can define a normal bundle $N(Y)$
of $Y$ with a projection
\begin{equation}\label{1.3}
 \pi :\ N(Y)\ \to\ Y\ .
\end{equation}
The metric on $X$ induces on $N(Y)$ a Riemannian metric
\begin{equation}\label{1.4}
 g_\ve\=\pi^*g^{}_Y + \ve^2g^{}_Z\ ,
\end{equation}
where $Z\cong\R^4$ is a typical fibre. In fact, the fibres are calibrated by a 4-form $Q_\ve$ dual to $\Sigma_\ve$.
The metric (\ref{1.4}) extends to a tubular neighborhood of $Y$ in $X$, and (\ref{1.1}) may be considered
on this subset of $X$. Anyway, it was shown~\cite{5, 9, 6}
that solutions of the instanton equation (\ref{1.1}) defined by the form $\Sigma_\ve$ on $(X, g_\ve )$ in the adiabatic limit
$\ve\to 0$ converge to sigma-model instantons describing a map from the $(d{-}4)$-dimensional submanifold $Y$ into the
hyper-K\"ahler moduli space of framed Yang-Mills instantons on fibres $\R^4$ of the normal bundle~$N(Y)$.

The submanifold $Y\hra X$ is calibrated by the $(d{-}4)$-form $\Sigma$ defining the instanton equation (\ref{1.1}).
However, on $X$ there may exist other $p$-forms $\vp$ and associated $\vp$-calibrated submanifolds $Y$ of dimension
$p\ne d{-}4$. In such a case one can define a different normal bundle (\ref{1.3}) with fibres $\R^{d-p}$ and deform the metric as in (\ref{1.4}).
However, this task is quite difficult technically and will be postponed for a future work. As a more simple task,
one may take a direct product manifold $X=Y{\times}Z$ with dim$_{\R}Y=p$ and dim$_{\R}Z=q=d{-}p$ with
a $p$-form $\vp=vol_Y$, or consider non-flat manifolds $Z$ and a $(d{-}4)$-form $\Sigma$ defining (\ref{1.1}).
In string theory dim$_{\R}X=10$, and calibrated submanifolds $Y$ are identified with worldvolumes of $p$-branes where $p$ varies from zero to ten.

In this short paper we explore the direct product case $X=Y{\times}Z$ with
dim$_{\R}Y=p\ne d{-}4$ for K\"ahler manifolds $X$ and the adiabatic limit of the Hermitian Yang-Mills equations on bundles
over $X$.
We will show that for even $p$ (and hence even $q$) the adiabatic limit of (\ref{1.1}) yields sigma-model  instanton equations
describing holomorphic  maps from $Y$ into the moduli space of Hermitian Yang-Mills instantons on $Z$. For odd $p$ and $q$ the
consideration is more involved, and we describe only the case $p{=}q{=}3$ in which we obtain maps from $Y$ into the moduli
space of flat connections on $Z$. For the purpose of this paper, this special case sufficiently illustrates the main features of the odd-dimensional cases.

\section{Moduli space of instantons in $d\ge 4$}

{\bf Bundles.}
Let $X$ be an oriented smooth manifold of dimension $d$, $G$ a semisimple  compact Lie
group, $\gfrak$ its Lie algebra, $P$ a principal $G$-bundle over $X$, $\Acal$ a connection 1-form on $P$ and
$\Fcal =\diff\Acal + \Acal\wedge\Acal$ its curvature. We consider also the bundle of groups Int$P = P\times_G G$
($G$ acts on itself by internal automorphisms: $h\mapsto ghg^{-1},\ h, g\in G$) associated with $P$, the bundle
of Lie algebras Ad$P=P\times_G \gfrak$ and a complex vector bundle $E=P\times_GV$, where $V$ is the space of some irreducible representation of $G$. All these
associated bundles inherit their connection $\Acal$ from $P$.

\medskip

\noindent {\bf Gauge transformations.}
We denote by $\mathbb{A}'$ the space of connections on $P$ and by $\Gcal'$ the
infinite-dimensional
group of gauge transformations (automorphisms of $P$ which induce the identity transformation of $X$),
\begin{equation}\label{2.1}
\Acal\ \mapsto\ \Acal^g\=g^{-1}\Acal g + g^{-1}\diff g\ ,
\end{equation}
which can be identified with the space of global sections of the bundle Int$P$. Correspondingly,
the infinitesimal action of $\Gcal'$ is defined by global sections  $\chi$ of the bundle Ad$P$,
\begin{equation}\label{2.2}
\Acal\ \mapsto\ \de_\chi\Acal\=\diff\chi + [\Acal , \chi ]\ =:\ D^{}_{\Acal}\chi
\end{equation}
with $\chi\in\, $Lie$\Gcal'=\Gamma (X,\, $Ad$P)$.

\medskip

\noindent {\bf Moduli space of connections.}
We restrict ourselves to the subspace $\mathbb{A}\subset\mathbb{A}'$
of irreducible connections and to the subgroup $\Gcal = \Gcal' / Z(\Gcal')$ of $\Gcal'$ which acts freely on $\mathbb{A}$.
Then the {\it moduli space\/} of irreducible connections on $P$ (and on $E$) is defined as the quotient $\mathbb{A}/\Gcal$.
We do not distinguish connections related by a gauge transformation. Classes of gauge equivalent
connections are points $[\Acal ]$ in $\mathbb{A}/\Gcal$.

\medskip

\noindent {\bf Metric on $\mathbb{A}/\Gcal$.}
Since $\mathbb{A}$ is an affine space, for each $\Acal\in\mathbb{A}$
we have a canonical identification between the tangent space $T_{\Acal} \mathbb{A}$ and the space $\Lambda^1(X,\,$Ad$P$)
of 1-forms on $X$ with values in the vector bundle Ad$P$. We consider $\gfrak$ as a matrix Lie algebra, with the metric
defined by the trace. The metrics on $X$ and on the Lie algebra $\gfrak$ induce an inner product on $\Lambda^1(X,\,$Ad$P$),
\begin{equation}\label{2.3}
 \langle\xi_1, \xi_2\rangle \= \int_X\tr\,(\xi_1\wedge *\xi_2)
 \qquad\textrm{for}\qquad \xi_1, \xi_2\in \Lambda^1(X,\textrm{Ad}P)\ .
\end{equation}
This inner product is transferred to $T_{\Acal} \mathbb{A}$ by the canonical
identification. It is invariant under the $\Gcal$-action on $\mathbb{A}$, whence
we get a metric (\ref{2.3}) on the moduli space $\mathbb{A}/\Gcal$.

\medskip

\noindent {\bf Instantons.}
Suppose there exists a $(d{-}4)$-form $\Sigma$ on $X$ which allows us to introduce the instanton
equation
\begin{equation}\label{2.4}
 *\Fcal + \Sigma\wedge\Fcal \=0
\end{equation}
discussed in Section 1.
We denote by $\Ncal\subset\mathbb{A}$ the space of irreducible connections subject to (\ref{2.4})
on the bundle $E\to X$. This space $\Ncal$ of instanton solutions on $X$ is a subspace of the
affine space $\mathbb{A}$, and we define the moduli space $\Mcal$ of instantons as the quotient space
\begin{equation}\label{2.5}
\Mcal \=\Ncal /\Gcal
 \end{equation}
together with a projection
\begin{equation}\label{2.6}
\pi :\ \Ncal \stackrel{\Gcal}\to\Mcal \ .
\end{equation}

According to the bundle structure (\ref{2.6}), at any point $\Acal\in \Ncal$,
the tangent bundle $T_{\Acal}\Ncal\to\Ncal$ splits into the direct sum
\begin{equation}\label{2.7}
T_{\Acal}\Ncal \=\pi^*T_{[\Acal ]}\Mcal\oplus T_{\Acal}\Gcal \ .
\end{equation}
In other words,
\begin{equation}\label{2.7a}
T_{\Acal}\Ncal \ni\ \tilde\xi \= \xi + D_\Acal\chi \qquad\textrm{with}\qquad
\xi\in \pi^*T_{[\Acal ]}\Mcal  \and  D_\Acal\chi\in T_{\Acal}\Gcal\ ,
\end{equation}
where  $\tilde\xi , \xi\in \Lambda^1(X,\,$Ad$P$) and $\chi\in \Lambda^0(X,\,$Ad$P)=\Gamma(X,\,$Ad$P$).
The choice of $\xi$ corresponds to a local fixing of a gauge.

\medskip

\noindent {\bf Metric on $\Mcal$}.
Denote by $\xi_\a$
a local basis of vector fields on $\Mcal$ (sections of the tangent bundle $T\Mcal$) with $\a = 1,\ldots,\textrm{dim}_\R\Mcal$.
Restricting the metric (\ref{2.3}) on $\mathbb{A}/\Gcal$ to the subspace $\Mcal$
provides a metric ${\mathbb G}=(G_{\a\b})$ on the instanton moduli space,
\begin{equation}\label{2.8}
 G_{\a\b}\=\int_X \tr\,(\xi_\a\wedge *\xi_\b)\ .
\end{equation}

\medskip

\noindent {\bf K\"ahler forms on $\Mcal$.}
If $X$ is K\"ahler with a complex structure $J$ and a K\"ahler form $\om (\cdot , \cdot)=g(J\cdot , \cdot )$, then
the K\"ahler 2-form $\O =(\O_{\a\b})$ on $\Mcal$ is given by
\begin{equation}\label{2.9}
 \O_{\a\b}\=- \int_X\tr\,(J\xi_\a\wedge *\xi_\b)\ .
\end{equation}
It is well known that the moduli space of framed instantons\footnote{Framed instantons
are instantons modulo gauge transformations which approach the identity at a fixed point.}
on a hyper-K\"ahler 4-manifold $X$ (with three integrable almost complex structures $J^i$) is hyper-K\"ahler,
with three K\"ahler forms
\begin{equation}\label{2.10}
 \O_{\a\b}^i\=-\int_X\tr\,(J^i\xi_\a\wedge *\xi_\b)\ .
\end{equation}

\section{Hermitian Yang-Mills equations}

{\bf Instanton equations.}
On any K\"ahler manifold $X$ of dimension $d=2n$ there exists an integrable  almost complex structure $J\in\,$End($TX$),
$J^2 =-$Id, and a K\"ahler (1,1)-form $\om(\cdot,\cdot)=g(J\cdot,\cdot)$ compatible with $J$.
The natural 4-form
\begin{equation}\label{3.1}
 Q\=\sfrac12\om\wedge\om
\end{equation}
and its dual $\Sigma =*Q$ allow one to formulate the instanton equation (\ref{2.4}) for a connection $\Acal$ on a complex
vector bundle $E$ over $X$ associated to the principal bundle $P(X,G)$. The fibres $\C^N$ of $E$ support an irreducible $G$-representation.
For simplicity, we have in mind the fundamental representation of SU$(N)$.
One can endow the bundle $E$ with a Hermitian metric and choose $\Acal$
to be compatible with the Hermitian structure on $E$.

The instanton equations in the form (\ref{2.4}) with $\Sigma=\sfrac12*(\om\wedge\om)$
may then be rewritten as the following pair of equations,
\begin{equation}\label{3.2}
\Fcal^{0,2}\=-(\Fcal^{2,0})^\+\=0
\end{equation}
and
\begin{equation}\label{3.3}
\om^{n-1}\wedge\Fcal \=0\qquad \Leftrightarrow\qquad
\om\lrc\Fcal \= \om^{\hat\mu\hat\nu}\Fcal_{\hat\mu\hat\nu}\=0\ ,
\end{equation}
where $\mh,\nh,\ldots=1,\ldots,2n$, and the notation $\om\lrc$ exploits the underlying Riemannian metric of $X$
for raising indices of $\om$. The equations  (\ref{3.2})-(\ref{3.3}) were introduced by Donaldson, Uhlenbeck and Yau~\cite{18}
 and are called the Hermitian Yang-Mills (HYM) equations.\footnote{
 Instead of (\ref{3.3}) one sometimes finds $\om\lrc\Fcal=\im\,\la\,\Id_E$ with $\la\in\R$.
 We take $\la=0$, i.e.\ assume $c_1(E)=0$,
 since one may always pass from a rank-$N$ bundle of non-zero degree to one of zero degree by considering
 $\tilde\Fcal = \Fcal -\sfrac{1}{N}(\tr\Fcal ) {\bf 1}_N$.}
The HYM equations have the following algebro-geometric interpretation.
Equation (\ref{3.2}) implies that the curvature $\Fcal=\diff\Acal +\Acal\wedge\Acal$ is of type (1,1)
with respect to $J$, whence the connection $\Acal$ defines a {\it holomorphic structure\/} on $E$.
Equation (\ref{3.3}) means that $E\to X$ is a {\it polystable\/} vector bundle.
The moduli space $\Mcal_X$ of
HYM connections on $E$, the metric $\Gbb =(G_{\a\b})$ and the K\"ahler form $\O =(\O_{\a\b})$ on $\Mcal_X$ are introduced as
described in Section 2 after specializing $X$ to be K\"ahler.

\medskip

\noindent {\bf Direct product of K\"ahler manifolds.}
The subject of this paper is the adiabatic limit of the HYM equations (\ref{3.2})-(\ref{3.3})
on a direct product
\begin{equation}\label{3.4}
X\=Y\times Z
\end{equation}
of K\"ahler manifolds $Y$ and $Z$. The dimensions $p$ and $q$ of $Y$ and $Z$ are even, and $p+q=2n$. Let
$\{e^a\}$ with $a=1,\ldots,p$ and $\{e^\mu\}$ with $\mu= p{+}1,\ldots,2n$ be local frames
for the cotangent bundles $T^*Y$ and $T^*Z$, respectively.
Then $\{e^\mh\}=\{e^a, e^\mu\}$ with $\mh= 1,\ldots,2n$ will be a local frame for the cotangent
bundle $T^*X=T^*Y\oplus T^*Z$. We introduce on $Y\times Z$ the metric
\begin{equation}\label{3.5}
 g\=g_Y + g_Z\= \de_{ab}\,e^a\otimes e^b + \de_{\mu\nu}\,e^\mu\otimes e^\nu \= \de_{\mh\nh}\,e^\mh\otimes e^\nh
\end{equation}
and an integrable almost complex structure
\begin{equation}\label{3.6}
J\=J_Y \oplus J_Z \in \mbox{End}(TY)\oplus \mbox{End}(TZ)\ ,\quad J^2_Y=-\Id_Y\and J^2_Z=-\Id_Z\ ,
\end{equation}
whose components are defined by $J_Ye^a = J^a_be^b$ and $J_Ze^\mu = J^\mu_\nu e^\nu$.
Likewise, the K\"ahler form $\om (\cdot , \cdot )= g (J  \cdot , \cdot )$ on $Y\times Z$ decomposes as
\begin{equation}\label{3.8}
\om = \om_Y +\om_Z
\end{equation}
with components $\om_Y=(\om_{ab})$ and $\om_Z=(\om_{\mu\nu})$.

\medskip

\noindent {\bf Splitting of the HYM equations.}
We introduce on $X=Y\times Z$ local coordinates $\{y^a\}$  and $\{z^\mu\}$ and choose
$e^a=\diff y^a ,\ e^\mu=\diff z^\mu$. Any connection on the bundle $E\to X$ is decomposed as
\begin{equation}\label{3.9}
\Acal \=\Acal_Y + \Acal_Z \= \Acal_a\diff y^a + \Acal_\mu\diff z^\mu\ ,
\end{equation}
where the components $\Acal_a$ and $\Acal_\mu$ depend on $(y,z)\in Y\times Z$.
The curvature $\Fcal$ of $\Acal$ has components $\Fcal_{ab}$ along $Y$, $\Fcal_{\mu\nu}$ along $Z$, and
$\Fcal_{a\mu}$ which we call ``mixed''.

Note that the holomorphicity conditions (\ref{3.2}) may be expressed through the projector
\begin{equation}\label{3.10}
\bar P=\sfrac12\, (\Id + \im J)\ , \qquad \bar P^2 =\bar P
\end{equation}
onto the (0,1)-part of the complexification of the cotangent bundle $T^*X=T^*Y\oplus T^*Z$ as
\begin{equation}\label{3.11}
\bar P \bar P\Fcal \= 0\ ,
\end{equation}
which in components reads
\begin{equation}\label{3.12}
\bigl(\de_\mh^\sh + \im J_\mh^\sh\bigr)\bigl(\de_\nh^\lh + \im J^\lh_\nh\bigr) \Fcal_{\sh\lh}\=0\ .
\end{equation}
From (\ref{3.6}) it follows that these equations split into three parts:
\begin{equation}\label{3.13}
\bigl(\de_a^c + \im J_a^c\bigr)\bigl(\de_b^d + \im J_b^d\bigr) \Fcal_{cd}\=0
\qquad\Leftrightarrow\qquad \Fcal_Y^{0,2}=0\ ,
\end{equation}
\begin{equation}\label{3.14}
\bigl(\de_\mu^\sigma + \im J_\mu^\sigma\bigr)\bigl(\de_\nu^\la + \im J_\nu^\lambda\bigr) \Fcal_{\sigma\lambda}\=0
\qquad\Leftrightarrow\qquad \Fcal_Z^{0,2}=0\ ,
\end{equation}
and
\begin{equation}\label{3.15}
\Fcal_{a\nu}J^\nu_\mu + J_a^c\Fcal_{c\mu}\=0\qquad\Leftrightarrow\qquad \Fcal_{a\mu}-J^c_aJ_\mu^\nu\Fcal_{c\nu}\=0\ .
\end{equation}
Finally, with the help of (\ref{3.8}) the stability equation (\ref{3.3}) takes the form
\begin{equation}\label{3.16}
\om_Y\lrc\Fcal_Y +  \om_Z\lrc\Fcal_Z \= \om^{ab}\Fcal_{ab}+\om^{\mu\nu}\Fcal_{\mu\nu} \=0\ .
\end{equation}

\section{Adiabatic limit of the HYM equations for even $p$ and $q$}

{\bf Moduli space $\Mcal_Z$.}
In order to investigate the adiabatic limit of (\ref{3.13})-(\ref{3.16}),
we introduce on $X=Y\times Z$ the deformed metric and K\"ahler form
\begin{equation}\label{4.1}
g_\ve \= g_Y+ \ve^2g_Z \und \om_\ve \= \om_Y + \ve^2\om_Z\ ,
\end{equation}
while the complex structure $J=J_Y\oplus J_Z$ does not depend on $\ve$ according to (\ref{3.6}).
Since $J_Y$ and $J_Z$ are untouched, (\ref{3.13})-(\ref{3.15}) keep their form
in the adiabatic limit $\ve\to 0$.
In particular, (\ref{3.13}) implies that $\Fcal_Y^{0,2}=0$, i.e.\ the bundle
$E\to Y\times Z$ is holomorphic along $Y$ for any $z\in Z$.\footnote{
We can always choose a gauge such that $\Acal_Y^{0,1}=0$
and locally $\Acal_Y^{1,0}=h^{-1}\pa_Y h$ for a $G$-valued function $h(y,z)$.}
On the other hand, (\ref{3.16}) for $\ve\to 0$ becomes
\begin{equation}\label{4.3}
\om_Z\lrc\Fcal_Z \= \om^{\mu\nu}\Fcal_{\mu\nu} \= 0\ ,
\end{equation}
which together with (\ref{3.14}) means that $\Acal_Z$ is a HYM connection
(framed instanton) on $Z$ for any given $y\in Y$. We denote the moduli space of such connections by
\begin{equation}\label{4.4}
\Mcal_Z \= \Ncal_Z/\Gcal_Z\ ,
\end{equation}
where $\Ncal_Z$ is the space of all instanton solutions on $Z$ for a fixed $y\in Y$,
and $\Gcal_Z$ consists of the elements of $\Gcal$ with the same fixed value of $y$.
We here suppress the $y$ dependence in our notation.
The moduli space $\Mcal_Z$ is a K\"ahler manifold
on which we introduce the metric $\Gbb$ and K\"ahler form $\O$ with components
\begin{equation}\label{4.5}
 G_{\a\b}  \=\int_Z \tr\,(\xi_\a\wedge *_Z\xi_\b ) \und
 \O_{\a\b} \= -\int_Z \tr\,(J_Z\xi_\a\wedge *_Z\xi_\b )
\end{equation}
similar to (\ref{2.8}) and (\ref{2.9}) but now with $\xi_\a\in\Lambda^1(Z,$ Ad$P$)
and the Hodge operator $\ast_Z$ defined on $Z$. Note that
for $\mathrm{dim}_\R Z=2$  the HYM equations (\ref{3.14}) and (\ref{4.3}) enforce $\Fcal_Z =0$,
i.e.\ $\Mcal_Z$ becomes the moduli space of flat connections on bundles $E(y)$ over a two-dimensional
Riemannian manifold $Z$.

\medskip

\noindent {\bf A map into $\Mcal_Z$.}
The bundle $E(y)$ is a HYM vector bundle over $Z$ for any $y\in Y$.
Letting the point $y$ vary, the connection $\Acal_Z=\Acal_\mu(y,z)\diff z^\mu$ on $E(y)$ defines a map
\begin{equation}\label{map}
\phi : \ Y\ \to\ \Mcal_Z \qquad\textrm{with}\qquad
\phi(y) \= \bigl\{\phi^\a(y)\bigr\}\ ,
\end{equation}
where $\phi^\a$ with $\a=1,\ldots,\textrm{dim}_\R\Mcal_Z$ are local coordinates on $\Mcal_Z$.
This map is constrained by our remaining set of equations,
namely (\ref{3.15}) for the mixed field-strength components
\begin{equation}\label{4.7}
\Fcal_{a\mu}\=\pa_a\Acal_\mu - \pa_\mu\Acal_a + [\Acal_a, \Acal_\mu ]\=\pa_a\Acal_\mu - D_\mu\Acal_a\ .
\end{equation}
Similarly to (\ref{2.7}) and (\ref{2.7a}),
$\pa_a\Acal_\mu$ decomposes into two parts,
\begin{equation}\label{4.8}
 T_{\Acal_Z}\Ncal_Z \= \pi^*T_{[\Acal_Z]}\Mcal_Z \oplus T_{\Acal_Z}\Gcal_Z
 \qquad\Leftrightarrow\qquad
 \pa_a\Acal_\mu \= (\pa_a\phi^\a )\xi_{\a\mu} + D_\mu\eps_a\ ,
\end{equation}
where $\{\xi_\a = \xi_{\a\mu}\diff z^\mu\}$ is a local basis of vector fields on $\Mcal_Z$.
Here, $\eps_a$ are $\gfrak$-valued gauge parameters which are determined by the gauge-fixing equations
\begin{equation}\label{4.10}
(\pa_a\phi^\a)\,g^{\mu\nu}D_\mu\xi_{\a\nu} \= 0 \qquad\Rightarrow\qquad
g^{\mu\nu}D_\mu D_\nu \eps_a \=  g^{\mu\nu}D_\mu\pa_a\Acal_\nu\ .
\end{equation}
Substituting (\ref{4.8}) into (\ref{4.7}), the mixed field-strength components simplify to
\begin{equation}\label{4.11}
\Fcal_{a\mu}\=(\pa_a\phi^\a)\,\xi_{\a\mu}- D_\mu (\Acal_a-\eps_a) \ .
\end{equation}
Inserting this expression into our remaining equations (\ref{3.15}), we obtain
\begin{equation}\label{4.12}
(\pa_a\phi^\a )\,\xi_{\a\mu} - J^c_aJ^\sigma_\mu  (\pa_c\phi^\a )\,\xi_{\a\sigma} \=
D_\mu (\Acal_a-\eps_a) - J^c_aJ^\sigma_\mu D_\sigma (\Acal_c - \eps_c)
\end{equation}
as a condition on the map~$\phi$.

\medskip

\noindent {\bf Sigma-model instantons.}
In order to better interpret the above equations,
we multiply both sides with $\diff z^\mu\wedge *_Z\xi_\b$, take the trace over $\gfrak$,
integrate over $Z$ and recognize the integrals in (\ref{4.5}).
The integral of the right-hand side of (\ref{4.12}) vanishes due to
(\ref{4.8})-(\ref{4.10}) (orthogonality of $\xi_\a\in T\Mcal_Z$ and $D\chi\in T\Gcal_Z$),
and we end up with
\begin{equation}\label{4.13}
(\pa_a\phi^\a) G_{\a\b} + J^c_a\,(\pa_c\phi^\a) \O_{\a\b} \=0\ .
\end{equation}

Inverting the moduli-space metric $G$ and introducing the almost complex structure $\Jcal$ on $\Mcal_Z$
via its components
\begin{equation}\label{4.14}
\Jcal^\a_\b\ :=\ \O_{\b\ga}G^{\ga\a}\ ,
\end{equation}
we rewrite (\ref{4.13}) as
\begin{equation}\label{4.15}
 \pa_a\phi^\a \= -J^c_a\,(\pa_c\phi^\b) \Jcal^\a_\b  \qquad\Leftrightarrow\qquad
 \diff\phi \= - \Jcal \circ \diff\phi\circ J\ .
\end{equation}
Using $J^a_cJ^c_b= -\de^a_b$ and $\Jcal^\a_\ga \Jcal^\ga_\b =-\de^\a_\b$,
alternative versions are
\begin{equation}\label{4.16}
(\pa_a\phi^\b) \Jcal^\a_\b - J^b_a\,(\pa_b\phi^\a) \= 0 \qquad\Leftrightarrow\qquad
\Jcal\circ\diff\phi \= \diff\phi\circ J
\end{equation}
and
\begin{equation}\label{4.17}
(\de^b_a + \im J^b_a )\,(\pa_b\phi^\b) (\de^\a_\b-\im\Jcal^\a_\b) \= 0 \qquad\Leftrightarrow\qquad
\Pcal\circ\diff\phi\circ\bar P \=0 \ ,
\end{equation}
with the obvious definition for $\Pcal$.

These equations mean that $\phi^1+\im\phi^2,\ \phi^3+\im\phi^4,\ \ldots$ are holomorphic functions
of complex coordinates on $Y$, i.e.\ $\p$ is a holomorphic map.
It is clear that our equations (\ref{4.17}) are BPS-type (instanton)
first-order equations for the sigma model on $Y$ with target space $\Mcal_Z$, whose field equations
define harmonic maps from $Y$ into $\Mcal_Z$.
For $\mathrm{dim}_\R Y=\mathrm{dim}_\R Z=2$ these equations have appeared in~\cite{27}
as the adiabatic limit of the HYM equations on the product of two Riemann surfaces.\footnote{
See also~\cite{ber} where this limit was discussed in the framework of topological Yang-Mills theories.}
Our (\ref{4.17}) generalize~\cite{27} to the case $\mathrm{dim}_\R Y>2$ and $\mathrm{dim}_\R Z\ge 2$.
From the implicit function theorem it follows that near every solution $\phi$ of (\ref{4.17})
there exists a solution $\Acal_\ve$ of the HYM equations (\ref{3.2})-(\ref{3.3}) for $\ve$ sufficiently small.
In other words, solutions of (\ref{4.17}) approximate solutions of the HYM equations on $X$.

\section{Adiabatic limit of gauge instantons for $p=q=3$}

If the K\"ahler manifold $X$ is a direct product of two {\it odd\/}-dimensional manifolds $Y$ and $Z$,
i.e.\ if $p=\mathrm{dim}_\R Y$ and $q=\mathrm{dim}_\R Z$ are both odd, then we may need to impose
conditions on the geometry of $Y$ and $Z$ for $X=Y\times Z$ to be K\"ahler. However, we are not aware
of these demands outside of special cases, such as products of tori. Therefore, we
restrict ourselves to tori $Y$ and $Z$ with $p=q=3$ since already this case illustrates essential
differences from the case of even $p$ and $q$. More general situations demand more effort and
will be considered elsewhere.

\medskip

\noindent {\bf Deformed structures.}
We consider the Calabi-Yau space
\begin{equation}
X\= Y\times Z\= T^3\times T^3_r\ ,
\end{equation}
where $T^3$ is a 3-torus and $T^3_r$ is another 3-torus,
with $r$ marked points (punctures). We endow $X$ with the deformed metric
\begin{equation}\label{5.1}
g^{}_\ve \=g_{T^3} + \ve^2g_{T^3_r} \=
e^1\otimes e^1 + e^2\otimes e^2 + e^3\otimes e^3 + \ve^2(e^4\otimes e^4 + e^5\otimes e^5 + e^6\otimes e^6)
\end{equation}
and choose the basis of (1,0)-forms as
\begin{equation}\label{5.2}
\th^1=e^1+\im\ve e^4\ ,\quad  \th^2=e^2+\im\ve e^5\and \th^3=e^3+\im\ve e^6
\end{equation}
with a real deformation parameter $\ve$.

The combined torus $T^3\times T^3_r$ supports an integrable almost complex structure $J$
satisfying $J\th^j=\im \th^j$ for $j=1,2,3$, which determines its components,
\begin{equation}\label{5.3}
Je^\mh = J^\mh_\nh e^\nh :\quad
J^1_4=J^2_5=J^3_6=-\ve \and J_1^4=J^5_2=J^6_3=\ve^{-1} \ .
\end{equation}
For the K\"ahler form $\om (\cdot , \cdot )=g(J\cdot , \cdot )$ the components are
\begin{equation}\label{5.4}
\om_{14}=\om_{25}=\om_{36}=\ve \und \om_{41}=\om_{52}=\om_{63}=-\ve\ .
\end{equation}

\medskip

\noindent {\bf Adiabatic limit for instantons.}
The HYM equations  (\ref{3.2}) and  (\ref{3.3})
on $T^3\times T^3_r$ with $J$ and $\om$ given by (\ref{5.3}) and  (\ref{5.4}) read
\begin{equation}
\begin{aligned} \label{5.5}
 \Fcal_{ab}+\im\Fcal_{a\mu}J^\mu_b + \im J^\mu_a\Fcal_{\mu b} - J^\mu_aJ^\nu_b\Fcal_{\mu \nu}&\=0\ , \\[4pt]
 \Fcal_{\mu\nu}+\,\im\Fcal_{\mu b}J_\nu^b +\,\im J_\mu^b\Fcal_{b\nu } -\,J_\mu^aJ_\nu^b\Fcal_{ab}&\=0\ , \\[4pt]
 \Fcal_{a\mu}+\,\im\Fcal_{a b}J_\mu^b +\,\im J^\nu_a\Fcal_{\nu\mu } - J^\nu_aJ_\mu^b\Fcal_{\nu b}&\=0\ ,
\end{aligned}
\end{equation}
with $a,b=1,2,3$ and $\mu,\nu=4,5,6$, as well as
\begin{equation}\label{5.8}
 \Fcal_{14}+\Fcal_{25} + \Fcal_{36}\=0\ .
\end{equation}
In the adiabatic limit $\ve\to 0$ the first two lines of (\ref{5.5}) reduce to
\begin{equation}\label{5.9}
 \Fcal_{45}\=\Fcal_{46}\=\Fcal_{56}\=0
\end{equation}
while the mixed-component part of (\ref{5.5}) together with (\ref{5.8}) produces
\begin{equation}\label{5.10}
 \Fcal_{16} - \Fcal_{34}\=0\ , \quad
 \Fcal_{35} - \Fcal_{26}\=0\ , \quad
 \Fcal_{24} - \Fcal_{15}\=0 \und
 \Fcal_{14} + \Fcal_{25} + \Fcal_{36}\=0\ .
\end{equation}
Recall that
\begin{equation}\label{5.11}
\Acal \= \Acal_Y + \Acal_Z \= \Acal_a(y,z)\diff y^a + \Acal_\mu (y,z)\diff z^\mu
\end{equation}
is a connection on a vector bundle $E$ over $X=T^3\times T^3_r$.
From (\ref{5.9}) we learn that $\Acal_Z$ is a flat connection on $Z=T^3_r$ for any $y\in Y=T^3$.
We denote by $\Ncal_Z$ the space of solutions to (\ref{5.9}) and by $\Mcal_Z$ the moduli space of all such connections.
From (\ref{5.10}) we see that in the adiabatic limit there are no restrictions on $\Acal_Y$, since
the components $\Acal_a$ and $\Fcal_{ab}$ no longer appear.

\medskip

\noindent {\bf Sigma-model equations.}
For the mixed components $\Fcal_{a\mu}$ of the field strength we have
\begin{equation}\label{5.12}
 \Fcal_{a\mu}\=\pa_a\Acal_\mu - D_\mu\Acal_a \= (\pa_a\phi^\a )\xi_{\a\mu}-D_\mu (\Acal_a-\eps_a)
\end{equation}
where, as in Section 4, we used for $\pa_a\Acal_\mu$ the decomposition formula  (\ref{4.8})
and introduced the map
\begin{equation}\label{map2}
\phi : \ T^3 \ \to\ \Mcal_{T^3_r} \qquad\textrm{with}\qquad
\phi(y) \= \bigl\{\phi^\a(y)\bigr\} \ ,
\end{equation}
where $\phi^\a$ with $\a=1,...,\dim_{\R}\Mcal_{T^3_r}$ are local coordinates on $\Mcal_{T^3_r}$.

Substituting (\ref{5.12}) into (\ref{5.10}), we obtain the equations
\begin{equation}\label{5.13}
\begin{aligned}
 (\pa_1\phi^\a)\,\xi_{\a 6} - (\pa_3\phi^\a)\,\xi_{\a 4} &\= D_6 (\Acal_1-\eps_1)- D_4 (\Acal_3-\eps_3)\ , \\[4pt]
 (\pa_3\phi^\a)\,\xi_{\a 5} - (\pa_2\phi^\a)\,\xi_{\a 6} &\= D_5 (\Acal_3-\eps_3)- D_6 (\Acal_2-\eps_2)\ , \\[4pt]
 (\pa_2\phi^\a)\,\xi_{\a 4} - (\pa_1\phi^\a)\,\xi_{\a 5} &\= D_4 (\Acal_2-\eps_2)- D_5 (\Acal_1-\eps_1)\
\end{aligned}
\end{equation}
and
\begin{equation}\label{5.16}
 (\pa_1\phi^\a)\,\xi_{\a 4} + (\pa_2\phi^\a)\,\xi_{\a 5}+ (\pa_3\phi^\a)\,\xi_{\a 6}\=
 D_4 (\Acal_1-\eps_1)+ D_5 (\Acal_2-\eps_2)+D_6 (\Acal_3-\eps_3)\ .
\end{equation}
Multiplying both sides with $\xi_{\b\mu}$ for $\mu =4,5,6$
and integrating $\tr\,(\xi_{\a\mu}\xi_{\b\nu})$ over $T^3_r$, the above four equations
yield the $3\,\textrm{dim}_\R\Mcal_{T^3_r}$ relations
\begin{equation}\label{5.17}
\pa_a\phi^\a + \pi_a{\,}^b_c\,(\pa_b\phi^\b)\,\Pi^c{\,}^\a_\b \= 0\ ,
\end{equation}
where
\begin{equation}\label{5.19}
 \pi_a{\,}^b_c\ :=\ \ve^b_{ac} \und
 \Pi^a{\,}^\a_\b\ :=\ \Pi^a_{\b\ga} G^{\ga\a}
\end{equation}
with
\begin{equation}\label{5.21}
G_{\a\b} \= \int_{T^3_r}\!\diff^3 z\ \de^{\mu\nu}\,\tr\,(\xi_{\a\mu}\xi_{\b\nu}) \und
\Pi^a_{\a\b} \= \int_{T^3_r}\!\diff^3 z\ \ve^{a+3\,\mu\nu}\,\tr\,(\xi_{\a\mu}\xi_{\b\nu})\ .
\end{equation}
The integrals of the right-hand sides of (\ref{5.13}) and (\ref{5.16}) vanish due to the 
orthogonality of $\xi_\a\in T\Mcal_{T^3_r}$ and $D_{\mu}\chi\in T{\cal G}_{T^3_r}$.

The (1,1) tensors $\pi_a=(\ve^b_{ac})$, $a=1,2,3$, on $T^3$ and the (1,1) tensors
$\Pi_a=(\de_{ab}\Pi^b{\,}^\a_\b)$ on $\Mcal_{T^3_r}$ satisfy the identities
\begin{equation}\label{5.23}
\pi^3_a + \pi_a \=0 \und \Pi^3_a + \Pi_a \=0\ ,
\end{equation}
i.e.\ they define three so-called $f$-structures~\cite{28} correspondingly on $T^3$ and on $\Mcal_{T^3_r}$.
To clarify their meaning we observe that (\ref{5.23}) defines orthogonal projectors
\begin{equation}\label{5.24}
 P_a\ :=\ - \pi^2_a \und P_a^{\bot}\ :=\ \unity_3 + \pi^2_a
\end{equation}
of rank two and rank one on $T^3$ and similarly orthogonal projectors
\begin{equation}\label{5.25}
 \Pcal_a\ :=\ - \Pi^2_a \und \Pcal_a^{\bot}\ :=\ \Id + \Pi^2_a
\end{equation}
on $\Mcal_{T^3_r}$, where $\Id$ is the identity tensor.
The tangent bundle $T(T^3)$ splits into eigenspaces of $P_a$,
\begin{equation}\label{5.26}
 T(T^3)\=T(T^2_a\times S^1_a)\=T(T^2_a)\oplus T(S^1_a)\=L_a\oplus N_a \quad\for a=1,2,3\ ,
\end{equation}
which defines on $T^3$ two distributions $L_a$ and $N_a$ of rank two and one, respectively,
and decomposes the 3-torus in three different ways.
Analogously, the projector $\Pcal_a$ yields a splitting
\begin{equation}\label{5.27}
 T(\Mcal_{T^3_r}) \= \Lcal_a\oplus\Ncal_a
\end{equation}
which is in fact induced by the factorization of $T^3_r$ into a two-dimensional torus and a circle.

Our equations (\ref{5.17}) look similar to the adiabatic form of  the
$G_2$-instanton equations (for a definition see e.g.~\cite{5, 6, 11, 13}) on the 7-manifold
\begin{equation}\label{5.33}
 X\=Y\times Z\=T^3\times Z \quad\with Z=T^4\ ,\quad K3 \or \R^4\ .
\end{equation}
In the adiabatic limit of $\ve\to 0$ with the deformed metric $g_\ve =g_Y + \ve^2 g_Z$
the $G_2$-instanton equations become
\begin{equation}\label{5.35}
\pa_a\phi^\a + \ve_{ac}^b\,(\pa_b\phi^\b)\,\Jcal^c{\,}^\a_\b \= 0\ .
\end{equation}
This looks similar to (\ref{5.17}) and features three complex structures
$\Jcal^c=(\Jcal^c{\,}^\a_\b)$ (instead of $f$-structures $\Pi^c$) on the hyper-K\"ahler moduli space $\Mcal_Z$
of framed Yang-Mills instantons on the hyper-K\"ahler 4-manifold $Z$. These equations
were discussed e.g.\ in~\cite{6,12} in the form of Fueter equations.
In the above case (\ref{5.33}) they define maps $\phi : T^3\to\Mcal_Z$
which are sigma-model instantons minimizing the standard sigma-model energy functional.

The moduli space of Yang-Mills instantons in (\ref{5.33})-(\ref{5.35}) has dimensionality divisible by four 
(a hyper-K\"ahler manifold), and it allows for three complex structures $\Jcal^a$. In distinction, 
the dimension of the moduli space of flat connections on 3-tori $T^3$ is a multiple of three~\cite{1a,2a,3a}
($r$ punctures add $3r$ real parameters to the above moduli), and the three $f$-structures $\Pi^a$ in (\ref{5.19}) 
play the role of degenerate complex structures on the moduli space $\Mcal_{T^3_r}$. 
Solutions of (\ref{5.17}) approximate solutions of the HYM equation on $X=T^3\times T^3_r$.

\bigskip

\noindent
{\bf Acknowledgements}

\noindent
This work was partially supported by the Deutsche Forschungsgemeinschaft grant LE 838/13.

\end{document}